\documentclass[aps,nofootinbib,showpacs]{revtex4}
\usepackage{amsfonts}
\usepackage{amsmath}
\usepackage{amssymb}
\usepackage{color}

\newcommand{\bra}[1]{\left \langle #1 \right \rvert}
\newcommand{\ket}[1]{\left \lvert #1 \right \rangle}

\begin{document}


\title{ Exploring the Tomlin-Varadarajan quantum constraints in $U(1)^3$ loop quantum gravity: solutions and the Minkowski theorem }

\author{Jerzy Lewandowski} \email{jerzy.lewandowski@fuw.edu.pl }
\author{Chun-Yen Lin} \email{chun-yen.lin@fuw.edu.pl }
\affiliation{Faculty of Physics, University of Warsaw, Pasteura 5, 02-093 Warsaw, Poland
}

\begin{abstract}
We explicitly solved the anomaly-free quantum constraints proposed by Tomlin and Varadarajan for the weak Euclidean model of canonical loop quantum gravity, in a large subspace of the model's kinematic Hilbert space which is the space of the charge network states. In doing so, we first identified the subspace on which each of the constraints acts converginly, and then by explicitly evaluating such actions we found the complete set of the solutions in the identified subspace. We showed that the space of solutions consists of two classes of states, with the first class having a property that involves the condition known from the Minkowski theorem on polyhedra, and the second class satisfying a weaker form of the spatial diffeomorphism invariance. 
\end{abstract}

\pacs{04.60.Pp, 04.60.Ds. }

\maketitle

\section{ Introduction}
Following the pioneer works\cite{paper00}\cite{paper0} by Laddha and Varadarajan toward a non-trivial and anomaly-free quantum constraint algebra for canonical quantum gravity, Tomlin and Varadarajan had constructed\cite{paper1}\cite{paper2} new scalar constraint operators for the theory of weak Euclidean quantum gravity introduced by Smolin\cite{Smolin}. 

As a legitimate model for such purpose, the theory is governed by its own analogous scalar $C[N]$ and momentum $D[\vec{M}]$ constraints, whose algebra is isomorphic to the one in canonical general relativity. Under the loop formulation\cite{LQGref1}\cite{LQGref2}\cite{LQGref3} in the Ashtekar variables, it is also in a close analogy with the full theory of loop quantum gravity. On the other hand, the weak gravity limit\cite{Smolin} leads to a simplified configuration space of the $U(1)^3$ gauge fields $A_a^{i=1,2,3}$ instead of the original $SU(2)$ Ashtekar gauge fields. Also, the Euclidean truncation makes the scalar constraints much simpler without the Lorentzian term. The canonical conjugate pairs are given by $( A_a^i, E_i^a)$, with the densitized frame variables $E_i^a$ as the momenta. The work by the authors focuses on the quantization of the outstanding Poisson bracket $\left\{C[N], C[M]\right\}$ in the algebra, which is the only one with a structure functional and posing the main obstacle for the quantization. Their approach is based on the observation that, with a shift vector $\vec{\mathcal{N}}_i=\vec{\mathcal{N}}_i[N,E]$ as a certain phase space function, the exceptional algebra takes the form $\left\{C[N], C[M]\right\}=-3\sum_i\left\{D[\vec{\mathcal{N}}_i], D[\vec{\mathcal{M}}_i]\right\}$; this equality is robust since it holds also for the full general relativity. Their proposal is to properly quantize both sides of this equality.  

In their new construction\cite{paper1}\cite{paper2}, Tomlin and Varadarajan use the standard loop representation\cite{LQGref1}\cite{LQGref2}\cite{LQGref3} in the $U^3(1)$ setting to provide the kinematic Hilbert space $\mathbb K$. Their new quantization scheme is then applied to the constraints using a non-standard regularization, and it remarkably leads to a non-trivial quantum representation of the outstanding Poisson bracket, supported in a certain domain ${\mathbb D}\subset \mathbb{K}^*$ as 
\begin{eqnarray}\label{rep}
\left[\hat{C}[N],\hat{C}[N']\right]\bigg|_{\mathbb D}=-3\sum_j\left[\hat{D}[\vec{\mathcal{N}}_{j}],\hat{D}[\vec{\mathcal{N}}'_{j}]\right]\bigg|_{\mathbb D}, 
\end{eqnarray}
where the operators $\hat{C}[N]$ and $\hat{D}[\vec{\mathcal{N}}_{j}]$ are both obtained using the new scheme, which in particular quantizes $\vec{\mathcal{N}}_i[N,E]$ into operators called the electric shift operators. It is in this sense that the model restricted to the space $\mathbb D$ is anomaly-free among the quantum scalar constraints. In the follow-up paper\cite{paper2}, the authors further showed that the constraint operators and $\mathbb D$ can be slightly modified while retaining the algebra, so that the construction becomes covariant under the spatial diffeomorphisms. Including the exponentiated momentum constraints through this way, the full non-trivial quantum constraint algebra of the model is then faithfully realized.

In order to unfold the full content of this construction, we put forth our observations on some important elements lying implicitly in the original works. Particularly, we spell out the explicit action of the individual constraint operators $\hat{C}[N]$ and $\hat{D}[\vec{\mathcal{N}}_{j}]$ involved in the commutators on their proper domains. As it turns out, a large domain for $\hat{C}[N]$ and $\hat{D}[\vec{\mathcal{N}}_{j}]$ can be constructed in $\mathbb{K}^*$ following the spirit of the original construction. This domain is an extension of ${\mathbb D}$, and we are able to prescribe its maximal subspace $\mathbb V$ of solutions to the quantum constraints. Remarkably, the solution space $\mathbb V$ demonstrates the quantum constraints’ preference for two types of structures mathematically significant to loop quantum gravity. Specifically, the first type of solutions satisfies a condition taking the form of the Minkowski theorem of polyhedra, and the second type satisfies a weaker form of the spatial diffeomorphism invariance. Lastly, through a restriction to ${\mathbb V} \cap {\mathbb D}$, our prescription gives the complete solutions in ${\mathbb D}$ in terms of these preferred structures.

\section{Weak Euclidean Gravity Constraints and Tomlin-Varadarajan Quantization}

\subsection{ Classical Setting}

The classical $U^3(1)$ model theory is obtained by taking the Smolin's limit\cite{Smolin} of canonical general relativity, namely taking the $G_N\to 0$ limit of the theory in the Ashtekar formalism\cite{LQGref1}\cite{LQGref2}\cite{LQGref3}, while using a real Barbero-Immirzi parameter $\gamma$ for the Ashtekar variables and ignoring the Lorentzian term in the scalar constraints. For simplicity, in this paper we assume the $\mathbb{R}^3$ topology for the spatial slice $\Sigma$, for which we will use a single coordinate patch $\sigma: p\in\Sigma \mapsto x(p)\in\mathbb{R}^3$.

In the Hamiltonian formulation, the action of the theory is
\begin{eqnarray} \label{action}
S[E,A,\Lambda,{\vec N},N]=\int dt \int_{\Sigma} d^3x \,E^a_i \dot{A}_a^i-G[\Lambda]-D[\vec{N}]-H[N]
\end{eqnarray}
where the phase space coordinates are given by the $U(1)^3$ gauge fields $\{A_a^{i=1,2,3}\}$, and their conjugate momenta $\{E^a_{i=1,2,3}\}$ as densitized triad fields, and the other fields $\{\Lambda,{\vec N}, N \}$ are Lagrangian multipliers. As in the original work, the speed of light will be set to unity, and $G_N$ will have a unit of $[\text{length}][\text{mass}]^{-1}$. The pair $(A,E)$ is related to the Ashtekar varables $(\mathcal{A}, E)$ for the full theory through the scaling $A=G^{-1}_N{\mathcal A} $, which under the limit $G_N\to 0$ effectively changes the local $SU(2)$ frame rotation symmetry into the local $U(1)^3$ symmetry. In the full theory with a chosen $\gamma$, the fields $E^a_{i}$ determine the spatial metric, and the fields $\mathcal{A}_a^i$ additionally determine the extrinsic curvature of the spatial slice. The different values of $\gamma$ simply correspond to the different pairs of Ashtekar variables related by canonical transformations. The only nontrivial Poisson brackets between the fields are given by $\{A^i_a(x), E^b_j(y)\}= \gamma \,\delta_{i,j} \,\delta_{a,b} \,\delta(x-y)$. Just as the case in the full theory, the Hamiltonian is the sum of the Gauss constraints $G[\Lambda]$, the momentum constraints $D[\vec{N}]$ and the scalar constraints $H[N]$, where
\begin{eqnarray}\label{constraints}
G[\Lambda]=\int d^3x \Lambda^i \partial_a E_i^a\\
D[\vec{N}]=\int d^3x N^a (E^b_iF^i_{ab}-A^i_a\partial_bE^a_i)\\
H[N]=\int d^3x \frac{1}{2}N\epsilon_{ijk}E^a_iE^b_jF^k_{ab} 
\end{eqnarray}
with 
\begin{eqnarray}
F^i_{ab}=\partial_aA^i_b-\partial_bA^i_a.
\end{eqnarray}
These define a first class constraint system isomorphic to the ADM algebra in the Ashtekar formulation, with the fields ${\vec N}$ and $N$ corresponding to respectively the shift vector and inverse densitized lapse function. In this algebra, the exceptional Poisson bracket is given by
\begin{eqnarray} \label{algebra}
\{H[N],H[M]\}=D[\vec{\omega}]+G[A \cdot \vec{\omega}] \,;\,\,
\omega^a\equiv E_i^aE_i^b(MN_{,b}-NM_{,b}). 
\end{eqnarray}
This is especially challenging for its quantum representation, due to the presence of the shift vector $\vec{\omega}[E, M,N]$ as a structure functional. Varadarajan and Tomlin \cite{paper1} successfully addressed the difficulty by changing the lapse $N$ to be of density weight of $-\frac{1}{3}$, with which the scalar constraints take the form
\begin{eqnarray}\label{C(N)}
C[N]=\frac{1}{2}\int d^3x N \,q^{-\frac{1}{3}}\,\epsilon_{ijk}E^a_iE^b_jF^k_{ab}\,\, ; \,\, q\equiv \text{det}(E).
\end{eqnarray}   
Here $\sqrt{q}$ is the spatial volume density. Also, they introduced an ``electric shift" vector fields $\vec{\mathcal N}_i[N,E]$ which is given by the $E$ and $N$ fields as 
\begin{eqnarray} \label{electricshift}
 {\mathcal N}^a_i[N,E] \equiv N\,E^a_i\, q^{-\frac{1}{3}}.
\end{eqnarray}     
In terms of the new form of the constraints and the electric shift vector fields, the exceptional Poisson bracket (\ref{algebra}) can be re-expressed as    
\begin{eqnarray} \label{exceptional algebra}
\left\{C[N], C[M]\right\}=-3\sum_i\left\{D[\,\vec{\mathcal N}_i\,]\,, \,\,D[\,\vec{\mathcal M}_i\,]\right\}
\end{eqnarray}
We will follow the original work and call $D[\vec{\mathcal N}_i]$, the momentum constraints smeared the electric shift vector fields, the electric momentum constraints. In the following, we briefly describe the proposed quantization scheme applied to both $C[N]$ and $D[\vec{\mathcal N}_i]$. For the full prescription and technical details, the readers may refer to the listed original works\cite{paper1}\cite{paper2}.

\subsection{$U(1)^3$ Holonomy-Flux Algebra and Charge Network States}

The kinematic Hilbert space of the model theory follows from the standard loop representation of loop quantum gravity\cite{LQGref1}\cite{LQGref2}\cite{LQGref3}. In this representation, the $A$ fields are described by their holonomies over arbitrary curves, and the $E$ fields by their fluxes over arbitrary surfaces.

In our case, the holonomy $h_{e,\vec{q}}[A]$ is defined with an oriented smooth curve $e\subset \Sigma$ called an edge, and also a set of integer charge values $\{q^j\}\equiv \vec{q}$. Explicitly, it is defined as 
\begin{eqnarray} \label{holonomy}
h_{e,\vec{q}}[A]\equiv e^{i\kappa\gamma q^j \int_e A_a^j dx^a},
\end{eqnarray}
where $\kappa$ is a constant of dimension of length times inverse mass. Given a closed, oriented graph $\alpha$ consisting of a set of edges $\{e_I\}$ meeting only at their end points called the vertices, we may assign $\{\vec{q}_I\}$ to each of the edges and thereby defining the graph holonomy $h_{\alpha, \{\vec{q}_I\}}$ as
\begin{eqnarray} \label{graph holonomy}
h_{\alpha, \{\vec{q}_I\}}[A]\equiv \prod_{I} h_{e_I,\vec{q}_I}[A].
\end{eqnarray}

A graph holonomy is local $U(1)^3$ invariant and thus a solution to the Gauss constraints, if and only if the full set of edges $\{e_{I_v}\}$ sharing any vertex $v\in \alpha$ always satisfy the charge neutrality 
\begin{eqnarray} \label{invariance}
\sum_{I_v} \text{sgn}_{I_v} q^i_{I_v}=0
\end{eqnarray}
for all $i$, where $\text{sgn}_{I_v}$ is a positive or negative sign if the edge $e_{I_v}$ is out- or in-going for $v$. We now define a locally $U(1)^3$ invariant charge network state, denoted as $c\equiv c(\alpha, \{\vec{q}_I\})$, to be a kinematic quantum state with a wave functional given by its associated graph holonomy satisfying (\ref{invariance}). This wave functional as a graph holonomy can be written more compactly as $h_c$, along with the expression
\begin{eqnarray} 
\label{charge network}
h_c\equiv \exp (\int_{\Sigma} d^3x \,c_i^a A^i_a),
\end{eqnarray}
where 
\begin{eqnarray} \label{charge network coord}
c_i^a(x;\alpha,\{\vec{q}_I\})\equiv \sum_{I}i\kappa\gamma q_I^i \int dt_I \,\delta(e_I(t_I), x)\,\dot{e}^a_I(t_I)
\end{eqnarray}
is called the charge network coordinate. Note that the labeling $(\alpha, \{\vec{q}_I\})$ to the charge network functionals of $A$ is not unique, since we can always artificially change $\alpha$ into $\alpha'$ by adding trivial vertices and edges; to avoid this redundancy we will always label a charge network state by the corresponding oriented graph with the minimal number of edges.

The $U(1)^3$ invariant flux variables for $E$ is defined over an oriented surface $S$, given by 
\begin{eqnarray} \label{Flux}
E_i(S)\equiv \int_S \,\epsilon_{abc}\,E^a_i\, d\sigma^{bc}
\end{eqnarray}
The only non-trivial holonomy-flux Poisson brackets are
\begin{eqnarray} \label{flux-holonomy bracket}
\{h_c, E_j(S)\}= i \frac{\kappa\gamma}{2} \sum_{e_I\subset \alpha} \epsilon(e_I,S) \,q_I^j \,h_c
\end{eqnarray}
where $\epsilon(e_I,S)$ is the sign of the relative orientation between the given $e_I$ and $S$ when the two intersect, and is zero otherwise.

Upon this setting, we now define the $U(1)^3$ invariant kinematic Hilbert space $\mathbb K\equiv Span\{\ket{c}\}$ to be spanned by the basis of all the distinct charge network states and equipped with the inner product
\begin{eqnarray} \label{inner prod}
\langle \, c\,|\,c'\rangle= \delta_{c,c'}.
\end{eqnarray}
In this space, a holonomy operator acts as a multiplicative operator
 \begin{eqnarray} \label{holo op}
\hat{h}_c\ket{c'}\equiv \ket{c'+c},
\end{eqnarray}
where $c+c'$ is a new spin charge network with a graph which is the minimal one containing both $\alpha$ and $\alpha'$, and with the charge assignment given by the sum of $q_I^j$ and ${q'}_{I'}^{j}$. A flux operator then acts as a differential operator
\begin{eqnarray} \label{flux op}
 \hat{E}_j(S) \ket{c}=  \frac{\kappa\gamma\hbar}{2} \sum_{e_I\subset \alpha}\epsilon(e_I,S) \,q_I^j \ket{c}.
\end{eqnarray}

In loop quantum gravity, the kinematic states are instead the spin network states defined with the $SU(2)$ holonomies. The corresponding holonomy and flux operators are the elementary building blocks for the description of the spatial quantum geometry\cite {LQGref1}\cite{LQGref2}\cite{LQGref3}. Also, in such a kinematic Hilbert space $\mathbb{K}_{LQG}$, the spatial diffeomorphisms are naturally represented by the diffeomorphic deformations of the graphs of the spin network states without changing the spin assignments. The inner product in $\mathbb{K}_{LQG}$ of the same form as (\ref{inner prod}) allows the group averaging procedure to yield a diffeomorphism invariant Hilbert space ${\mathbb K}_{LQG}^{\mathit{diff}}\subset {\mathbb K}_{LQG}^*$, which solves the exponentiated momentum constraints. The space ${\mathbb K}_{LQG}^{\mathit{diff}}$ is simply spanned by a basis consisting of the dual states labeled by the distinct diffeomorphism classes of the spin network states. The quantization of the remaining scalar constraints is far more intricate in this procedure, and must rely on a special regularization in order to use the loop variables. All of these features are shared by our model in its kinematics given above, upon which the Tomlin-Varadarajan loop quantization scheme was applied for the quantum scalar constraints.

\subsection{ Tomlin-Varadarajan Loop Quantization of the Scalar Constraints}

 A general loop quantization procedure for any classical phase space function $\mathcal{F}[A,E]$ is through specifying the action of the corresponding operator on any spin (charge) network state in $\mathbb K$, and it mainly contains the following two steps. The first step is a regularization of $\mathcal{F}[A,E]$ with a suitable partition of the space $\Sigma$. In the loop representation, the elementary operators in $\mathbb K$ are the holonomy and the flux operators defined with the edges and surfaces of finite sizes, so we need a partition whose cells specify a set of small edges and surfaces, such that an approximant of $\mathcal{F}$ can be obtained in terms of the corresponding holonomy and flux variables. To an arbitrary spin (charge) network state $c$, a proper scheme then assigns one such partition consistent with the graph of the state, with the cell sizes controlled by a regulator parameter $\delta$. This leads to the approximant $\mathcal{F}^\delta_c$ satisfying $\lim_{\delta \to 0}\mathcal{F}^\delta_c= \mathcal{F}$, which is a function of the holonomy and flux variables associated to the partition consistent with $c$. The second step is the canonical quantization through promoting the classical loop variables into the corresponding operators defined in \eqref{holo op} and \eqref{flux op}, and this results to an operator $\mathcal{F}^\delta_c$. Finally, the desired operator is defined as {$\hat{\mathcal{F}}\equiv \lim_{\delta\to 0}\hat{\mathcal{F}}^{\delta} \equiv \lim_{\delta\to 0}\sum_{c} \hat{\mathcal{F}}^\delta_c\ket{c}\bra{c}$}, where the sum is taken over the orthonormal basis of the spin (charge) network states.

For an $\mathcal{F}$ given by an integral over $\Sigma$, $\hat{\mathcal{F}}_c^{\delta}$ is generally a sum over terms each corresponding to a cell of the partition. Among these terms, the ones with no corresponding intersection with the graph of $c$ would simply annihilate $c$ with their factors of flux operators. The terms with corresponding intersections with the graph would not only change the vertex configuration of $c$ with their factors of flux operators, but also change the graph and the spin(charge) of $c$ with their factors of holonomy operators. In our $U(1)^3$ case, these operations are just given by  \eqref{flux op} and \eqref{holo op}. Here, we will follow \cite{paper1,paper2} for the Tomlin-Varadarajan construction applied to the cases with $\mathcal{F}= C[N]$ and $\mathcal{F}'= D[\,\vec{\mathcal{N}}_i\,]$, which are involved in the outstanding bracket (\ref{exceptional algebra}). 

The standard schemes of the scalar constraint quantization\cite{LQGconstr0}\cite {LQGref1}\cite{LQGref2}\cite{LQGref3} result to a graph-changing operator $\hat{C}^{\delta}[N]$, which adds small loops to a spin(charge) network state at its vertices. These small loops represent the curvature factors in $C[N]$. Since these small loops shrink with the vanishing limit of $\delta$, according to \eqref{inner prod} the resulted sequence of states in taking the limit do not converge in $\mathbb K$. In these common schemes, $\hat{C}[N]$ is instead defined in some special subspace of the dual space ${\mathbb K}^*$, where the small loops in the shrinking sequence are effectively indistinguishable from one another. A paradigmatic example is the subspace ${\mathbb K}^*_{\mathit{diff}}$ of the diffeomorphism invariant states which ignore the exact sizes of the small loops. Moreover, it is shown that the diffeomorphism invariance may be relaxed to a partial diffeomorphism invariance \cite{habitatLewandowskiSahlmann} for the convergence.           

Naturally, the features of these constraint operators greatly depend on the chosen quantization schemes. The standard schemes described above lead to a well-defined operator on the suitable subspaces of ${\mathbb K}^*$, but they face the challenge of providing a well-defined quantum constraint algebra that is anomaly free. In the improved standard quantization schemes \cite{LQGconstr1}\cite{LQGconstr2}\cite{LQGconstr3} recently proposed, the scalar constraint operators commutators can be defined in a partially diffeomorphism invariant subspace of ${\mathbb K}^*$, called the ``vertex Hilbert space". However, these commutators are shown to be identically zero, giving $[\hat{C}[N], \hat{C}[M]]=0$ in the vertex Hilbert space that is not fully diffeomorphism invariant. This undesirable result is known as due to the ultra-locality of the constraint operators, which is the fact that the small loops added to two neighboring vertices of a graph are not coupled to one another. 

For the model with the non-ultralocal quantum constraint algebra \eqref{rep}, faithfully representing \eqref{exceptional algebra}, Tomlin and Varadarajan introduced an alternative loop regularization motivated by the following observations.

(i) The classical scalar constraints can be rewritten as 
\begin{eqnarray} \label{4.11}
 C[N]=-\frac{1}{2}\int_\Sigma d^3x \epsilon^{ijk} ({\mathcal{L}}_{\vec{\mathcal{N}}_j} A^k_b)E_i^b,=-\frac{1}{2}\int_\Sigma d^3x \,N\, \epsilon^{ijk} ({\mathcal{L}}_{\vec{\mathbf{N}}_j} A^k_b)E_i^b\,\,;\,\, {\mathbf{N}}_j[E] \equiv {\mathcal{N}}_j[1,E]
\end{eqnarray}
where we have extracted the lapse function as a prefactor, so that the final Lie derivatives are taken using the electric shift vectors \eqref{electricshift} over the lapse function. Also, we have ignored a term proportional to the Gauss constraints that is already solved in $\mathbb K$. Similarly, the electric momentum constraints can be expressed (again ignoring the Gauss constraint term) as
\begin{eqnarray} \label{5.7}
D[\vec{\mathcal{N}}_i]=\int_\Sigma d^3x ({\mathcal{L}}_{\vec{\mathcal N}_i} A^j_b)E_j^b=\int_\Sigma d^3x\, N\,({\mathcal{L}}_{{\bold{N}}_i} A^j_b)E_j^b.
\end{eqnarray}
This formulation suggests that the actions of both the constraints involve Lie dragging the $A$ fields in a set of directions ${\bold{N}}_j[E]$ determined by the $E$ fields.

(ii) A loop quantization of \eqref{4.11} and \eqref{5.7} {\color{blue}} involves replacing ${\mathbf{N}}_j[E]$ with certain quantum operators ${\mathbf{N}}_j$ constructed from the loop operators. According to (\ref{electricshift}), ${\mathbf{N}}_j$ will be a product between an inverse volume operator (to the power of $2/3$) built from the flux operators and another flux operator carrying the charge label $j$. As a well-established feature of the spatial volume operators in loop quantum gravity, which is obtianed from setting $\mathcal{F}= \int_{R\subset\Sigma}\, \sqrt{q}d^3$ in the previous section, the quanta of the volume are carried by the vertices of a charge network state. Also, the inverse volume operator is obtianed from setting $\mathcal{F}= \int_{R\subset\Sigma}  q^{-1/2}\,d^3x$, and again with the inverse volume quanta carried by the vertices. The loop variable corrections in these operators dominate in the small volume region, such that the vertices carrying zero spatial volume also carries zero (rather than infinite) inverse volume. Such a vertex with zero volume is called a degenerate vertex, and hence ${\mathbf{N}}_j$ acts non-trivially only on a non-degenerate vertex of a charge network state. When that happens, the flux operator labeled by $j$ receives individual contributions from all the edges connected to such a vertex. For a charge network state acted upon by the quantum constraints, we thus expect the quantum electric shift $N\cdot{\mathbf{N}}_j$ to be strongly peaked with non-zero values only at the non-degenerate vertices; at each of these vertices the quantum electric shift should be a sum over the edge-wise contributions from all the edges connected to the vertex.
 
(iii) A proper loop quantization of \eqref{4.11} and \eqref{5.7} can be introduced by approximating each of the Lie derivatives with a finite difference caused by a small Lie dragging parametrized by the regulator $\delta$. Specifically, since the $A$ fields are described by the holonomies, each of the Lie derivatives in the final forms of \eqref{4.11} and \eqref{5.7} should be approximated by a product, taken between the inverse of the corresponding holonomy and the image of this holonomy under the Lie dragging parametrized by the $\delta$ and generated by ${\mathbf{N}}_j$.

Based on the observations (i) and (ii), we expect the quantization of \eqref{4.11} and \eqref{5.7} to give
\begin{eqnarray} \label{sum form}
\hat{C}[N]=\lim_{\delta \to 0}\hat{C}^\delta[N]=\lim_{\delta \to 0} \sum_{x\in \Sigma}\, N_{x} (\hat{C}_{x})^\delta=\lim_{\delta \to 0} \sum_{x\in \Sigma}\sum_c\, N_{x} (\hat{C}_{x})^\delta_c \ket{c}\bra{c}\,\, \text{and}\nonumber 
\\
\hat{D}[\vec{\mathcal{N}}_j]=\lim_{\delta \to 0}\hat{D}^\delta[\vec{\mathcal{N}}_j]=\lim_{\delta \to 0} \sum_{x}\, N_{x} (\hat{D}^j_{x})^\delta=\lim_{\delta \to 0} \sum_{x\in \Sigma}\sum_c\, N_{x} (\hat{D}^j_{x})^\delta_c \ket{c}\bra{c}\,,
\end{eqnarray}
where the operators $(\hat{C}_{x})^\delta_c$ and $(\hat{D}^j_{x})^\delta_c$ can be nonzero only when $x$ coincides with the location of a non-degenerate vertex of $c$. Suppose $x_1$ coordinatize a nondegenerate vertex of $c$, then the operators $(\hat{C}_{x_1})^\delta_c$ and $(\hat{D}^j_{x_1})^\delta_c$ then act on this vertex with the operations encoding the corresponding Lie derivatives generated by ${\mathbf{N}}_j$ around $x_1$. Further, we note that $c$ is by construction an eigenstate of all the flux operators and hence of ${\mathbf{N}}_j$, and so each $c$ is assigned with a certain value of ${\mathbf{N}}_j$. Based on (ii), the value of ${\mathbf{N}}_j$ should be zero outside of the small balls $\{B_{ x_i, \delta}\}$ of radius $\delta$ around the locations of the vertices $\{ x_i \}$, and it should take the form of ${\mathbf{N}}_j\equiv \sum_{x_i}\sum_{I_{x_i}}({\mathbf{N}}_j)_{I_{x_i}}$ with the edge-wise contributions $({\mathbf{N}}_j)_{I_{x_i}}$ at each $x_i$. Recall that ${\mathbf{N}}_j$ contains the inverse volume eigenvalues carried by the by the vertices of $c$ which we will denote as $\{s_{x_i}\}$. Also, the value $s_{x_i}= s_{x_i}(q^j_{I_{x_i}})$ is determined by the neighboring charges of the vertex $x_i$ as detailed in the appendix of \cite{paper1}, such that $s_{x_i}=0$ if the vertex is trivial with zero volume.  Altogether, we then expect $({\mathbf{N}}_j)_{I_{x_i}}= {s_{x_i}}^{2/3} \,q^j_{I_{x_i}}\vec{e}_{I_{x_i}}$, where with all the charge dependence extracted $\vec{e}_{I_{x_i}}$ depends only on the graph $\alpha$ of $c$.

 The actions of $(\hat{C}_{x_i})^\delta_c$ and $(\hat{D}^j_{x_i})^\delta_c$ then change the state $c$ only in the region $B_{ x_i, \delta}\cap \alpha$. Bases on (iii), the Lie derivatives in $(\hat{D}^j_{x_i})^\delta_c$ can be represented by the product between 
\begin{eqnarray} \label{charge network B} 
h^{-1}_{c;x_i,\delta}\equiv \exp (-\int_{B_{ x_i,\delta}} d^3x \,c_i^a A^i_a),
\end{eqnarray}
and its Lie dragged images ${\mathcal T}^\delta_{\vec{e}_{I_{x_i}}} h_{c;x_i,\delta}$. This way, $(\hat{D}^j_{x_i})^\delta_c$ represents the Lie derivatives in the form
\begin{eqnarray} \label{D form}
 \delta^{-1}\sum_{I_{x_i}} \left(\Omega_{[x_i,I,\delta]}-\hat{I} \right)\,\,;\,\,\,\Omega_{[x_i,I,\delta]}\,\equiv\, h_{c;x_i,\delta}^{-1} \cdot {\mathcal T}^\delta_{\vec{e}_{I_{x_i}}}h_{c;x_i,\delta}.
\end{eqnarray}
Since $\vec{e}_{I_{x_i}}$ decays to zero outside of $B_{x_i,\delta}$, the graph holonomy $\Omega_{[x_i,I,\delta]}$ is based on a closed graph and is itself a gauge invariant charge network. The new state $\Omega_{[x_i,I,\delta]}\ket{c}$ has a clear structure. The term $h_{c;x_i,\delta}^{-1}$ erases the $h_{c;x_i,\delta}$ part of the charge network $c$, which is then replaced with the newly created part ${\mathcal T}^\delta_{\vec{e}_{I_{x_i}}}h_{c;x_i,\delta}$. The operator $\Omega_{[x_i,I,\delta]}$ thus deforms $c$ in a neighborhood of $x_i$, by moving the vertex $x_i$ in the direction of $\vec{e}_{I_{x_i}}$ together with segments of the attached edges contained in the neighborhood $B_{\delta;x_i}$. This deformation is diffeomorphic except for creating kinks at the points $\alpha \cap \partial B_{\delta;x_i}$, and these kinks become new vertices created by the operation. Overall, the $\Omega_{[x_i,I,\delta]}$ erases the old vertex at $x_i$, while creating the new ones located at $\alpha \cap \partial B_{\delta;x_i}$ and ${\mathcal T}^\delta_{\vec{e}_{I_{x_i}}}\, x_i$. We will figuratively refer to the last of the created vertices as the ``apex vertex" of $\Omega_{[x_i,I,\delta]}$.

Similarly, the Lie derivatives in \eqref{4.11} can be represented in $(\hat{C}_{x_i})^\delta_c$ through
\begin{eqnarray} \label{D form}
 \delta^{-1}\sum_{I_{x_i}} \left(\Delta_{[x_i,I,j,\delta]}-\hat{I} \right)\,\,;\,\,\,\Delta_{[x_i,I,j,\delta]}\,\equiv\, {\eta}_j\cdot\Omega_{[x_i,I,\delta]},
\end{eqnarray}
where an additional charge flipping $\eta_j$ operation appears due to the factor $\epsilon^{ijk}$ in the classical integrand. The charge flipping operation ${\eta}_j$ is associated to the $j$th charge and acts on the charges $q^1,q^2,q^3$ at every edge of the charge network $\Omega_{[x_i,I,\delta]}$ in the following way
\begin{eqnarray}
\eta_j(q^i)\ =\ q^j-\epsilon^{jki}q^k.
\end{eqnarray}
The resulting graph holonomy $\Delta_{[x_i,I,j,\delta]}$ is related to $\Omega_{[x_i,I,\delta]}$ by the charge flipping ${\eta}_j$ and is again gauge invariant. Due to this charge flipping $\Delta_{[x_i,I,j,\delta]}$ acts in a more complicated way. Instead of completely erasing $h_{c;x_i,\delta}$, the factor ${\eta}_j\cdot h_{c,x_i,\delta}^{-1}$ recharges it, and so the original graph vertex $x_i$ remains. Also, similar to the previous case, a new part given by ${\eta}_j\cdot{\mathcal{T}}^\delta_{\vec{e}_{I_{x_i}}}h_{c,x_i,\delta}$ is created. Overall, the $\Delta_{[x_i,I,j,\delta]}$ keeps the old vertex $x_i$ and creates the new ones at $\alpha \cap \partial B_{\delta;x_i}$ and also at the ``apex" ${\mathcal T}^\delta_{\vec{e}_{I_{x_i}}}\, x_i$.

The original work \cite{paper1} provides a concrete regularization scheme realizing all of the above expectations. Particularly, for each charge network state $c$ and a small enough $\delta$, the corresponding $\Delta_{[x_i,I,j,\delta]}$ and $\Omega_{[x_i,I,\delta]}$ are uniquely specified. The specified $\vec{e}_{I_{x_i}}$ is given by a vector field tangent to the edge ${e}_{I_{x_i}}$ at $x_i$, oriented as the edge were incoming to $x_i$, and normalized at $x_i$ with respect to the coordinates (in a covariant way as in \cite{paper2}). Just as in the standard scheme, the regulator parametrizes diffeormorphims on the graph structure. That is, for small enough $\{\delta,\delta'\}$, the states $\Delta_{[x_i,I,j,\delta]} \ket{c}$ and $\Omega_{[x_i,I,\delta]} \ket{c}$ are respectively diffeomorphic to $\Delta_{[x_i,I,j,\delta']} \ket{c}$ and $\Omega_{[x_i,I,\delta']} \ket{c}$. On the other hand, the created kinks mentioned above are such that the different states $\Delta_{[x_i,I,j,\delta]} \ket{c}$ or $\Omega_{[x_i,I,\delta]} \ket{c}$ associated with different edge lables $I$ are diffeomorphically inequivalent. Finally, the resulted regularized operators are given by 
\begin{eqnarray} \label{action 00}
(\hat{C}_{x})^\delta_c=  (s_{x_i})^{\frac{2}{3}}\, \sum_{I_{x_i},j} q^j_{I_{x_i}}\, \delta^{-1} \,\left(\Delta_{[x_i,I,j,\delta]}-\hat{I} \right) \,;\,\,(\hat{D}^j_{x_i})^\delta_c= (s_{x_i})^{\frac{2}{3}}\, \sum_{I_{x_i}} q^{j}_{I_{x_i}}\, \delta^{-1} \,\left(\Omega_{[x_i,I,\delta]}-\hat{I} \right).
\end{eqnarray}
The main result in the paper \cite{paper1} is the existence of the domain $\mathbb{D}$ in ${\mathbb K}^*$ where the commutators between $\hat{C}[N]=\lim_{\delta \to 0}\hat{C}^{\delta}[N]$ and $\hat{D}[\vec{\mathcal{N}}_j]=\lim_{\delta \to 0} \hat{D}^{\delta}[\vec{\mathcal{N}}_j]$ converge and represent \eqref{exceptional algebra} nontrivially. 

A crucial difference between this new scheme and the standard one lies in its representation of the curvature. Instead of the multiplication of a small holonomy loop based at the original vertex acted upon, the new operators act to move the location of the non-degenerate vertices. As shown in the original work, the operators always leave the original vertex $x_i$ acted upon degenerate, while supplying a new non-degenerate one which is the apex of either $\Delta_{[I_{x_i},j,\delta]}$ or $\Omega_{[I_{x_i},\delta]}$. Another key difference is the explicit presence of the $\frac{1}{\delta}$ factor. The standard LQG regularization takes advantage of the possibility of absorbing all the powers of $\delta$ into a finitely definite quantum operators. All of these features involve a chosen set of coordinate patches $\{\sigma_{c,i}\}$ appearing explicitly in the definition of the operator $\hat{C}[N]$, with each of the members covering one vertex neighborhood $i$ of a state $\ket{c}$. The covariance of $\hat{C}[N]$ under the spatial diffeomorphisms then demands a proper choice of such a set for the complete charge network basis. This is a technical problem carefully addressed in \cite{paper2}, and such a choice is shown to be given by consistently assigning the coordinate patches to the members of each diffeomorphism class of the charge network states, so that the coordinates patches for any two diffeomorphic states are also related by the same spatial diffeomorphisms, up to some irrelevant state preserving diffeomorphisms. Together with some mild modifications of the construction not affecting its key results above, the operator $\hat{C}[N]$ defined with this set of coordinate patches transforms correctly under the spatial diffeomorphisms with the correct density weight of $N$. 

In the first parts of our analysis we will calculate the single action by $\hat{C}[N]$ or $\hat{D}[\vec{\mathcal{N}}_j]$, using still the simplest (non-covariant) choice with $\{\sigma_{c,i}\equiv \sigma\}$. Nevertheless, the generality of the result will allow us to switch to the covariant construction when finding the solutions, for which we expect the general covariance becomes important.  We will carefully study the removal of the regulator to find the suitable domains for the quantum constraints. Then, we will explicitly solve the quantum constraints in such domains, and thereby extracting the remarkable features of the physical states for the model.

\section{Action, Domains, and Solution Space of $\hat{C}[N]$ }

\subsection{The action and the dual action}

The total quantum scalar constraint is given by $\hat{C}[{N}]=\sum_{x\in \Sigma}\, N(x)\ \lim_{\delta \to 0} (\hat{C}_x)^{\delta}$, where the operator $(\hat{C}_x)^{\delta}$ is defined by its action on the charge network basis as
\begin{eqnarray} \label{action}
(\hat{C}_{x'})^{\delta}\ket{c} = \sum_{x_i} \,\delta_{x',x_i}\,\, (s_{x_i})^{\frac{2}{3}}\, \sum_{I_{x_i},j} q^j_{I_{x_i}}\, \delta^{-1} \,\left(\Delta_{[x_i,I,j,\delta]}-\hat{I} \right) \ket{c},
\end{eqnarray}
The part of the sum defined by the term $-\hat{I}$ in the parentheses vanishes, due to the Gauss constraint (\ref{invariance}), therefore
\begin{eqnarray} \label{action 0}
(\hat{C}_{x'})^{\delta}\ket{c} = \sum_{x_i} \,\delta_{x',x_i}\,\, (s_{x_i})^{\frac{2}{3}}\, \sum_{I_{x_i},j} q^j_{I_{x_i}}\, \delta^{-1} \,\Delta_{[x_i,I,j,\delta]}\ket{c}.
\end{eqnarray}
According to \eqref{inner prod}, the general dual action of $\Delta_{[x',I,j,\delta]}$ removes $\Delta_{[x',I,j,\delta]}$ from a dual state $\bra{c'}$ whenever $c'$ can be written as
\begin{eqnarray}\label{decompdelta}
c'\ = \Delta_{[x_{i},I,j,\delta]} \cdot c \equiv  c_{x_{i},I,j,\delta} 
\end{eqnarray}
for a certain $x_{i}$ of $c$ with $x_{i}=x'$; otherwise, it simply annihilates $\bra{c'}$.  Fortunately, for every charged network $c'$, a decomposition (\ref{decompdelta}) with any given $(x_{i},I,j)$ is either unique or absent. Therefore, the dual operation of $\Delta_{[x_{i},I,j,\delta]}$ is given by 
\begin{eqnarray}
\label{factoring}
\Delta_{[x_{i},I,j,\delta]}: \bra{c_{x_i,I,j,\delta}}\ \mapsto\  \bra{c} \,\, \text{, and}\,\,\,\Delta_{[x_{i},I,j,\delta]}: \bra{c'}\ \mapsto \ 0 
\end{eqnarray}
for all other cases.

\subsection{The Habitat $T'_{*}$} 

Given a charge network $c$ with its ordered set of vertices $\{x_1, ... ,x_m\}\subset \Sigma$, and a smooth function $f\in C^\infty(\Sigma^m)$, we define a dual state 
\begin{eqnarray} \label{basis state 0}
\bra{c, f}\equiv \sum_{\phi\in \mathit{Diff}_{c}} f(\phi(x_1),...,\phi(x_n)) \bra{\, c\cdot \phi^{-1}}.
\end{eqnarray}
Here the sum ranges over a maximal subset $\mathit{Diff}_{c}$ of the diffeomorphism group, whose elements map $\ket{c}$ to all the distinct images  in $\mathbb{K}$ without redundancy. When there is a network symmetry in $c$ corresponding to a nontrivial diffeomorphism $\phi_s$ satisfying $c\cdot \phi_s^{-1}=c$, we further impose the consistency condition for the above, demanding that 
\begin{eqnarray}\label{consf}f(x_1,...,x_n)\ =
 f(\phi_s(x_1),...,\phi_s(x_n)).
 \end{eqnarray}
The description of the state $\bra{c, f}$ depends on the ordering of the vertices, however any permutation $\pi$ is equivalent to using the old ordering and replacing the function $f$ with $f\circ\pi $. Also, it is not hard to see that for every diffeomorphism $\phi$
\begin{eqnarray} \label{cphi}
\bra{c\circ \phi^{-1}, f}\ =\ \bra{c, f} .
\end{eqnarray}

The so-called habitat space proposed in \cite{LewMar} is spanned by the states of the form (\ref{basis state 0}) and is given by
\begin{eqnarray} \label{Habitat}
T'_{*}\equiv \mathit{\rm Span}\left\{\bra{c, f}\ :\ c \ \ {\rm is\ \ a\ \ charge\ \ network}, \ \ \ f\in C^\infty(\Sigma^m(c))\right\},
\end{eqnarray}
where $m(c)$ stands for the number of vertices of $c$. To construct a maximal set of linearly independent states, for each diffeomorphism class $[c]$ of a charge network $c$ choose an ordering of the vertices modulo the possible symmetries of $c$. That is, if $c$ has a non-trivial symmetry group, choose a class of orderings of the vertices of $c$, an orbit of the action of the symmetry group. The set of all the pairs $([c],f)$, where  $f\in C^\infty(\Sigma^m)$ such that (\ref{consf}) gives rise via 
(\ref{basis state 0}) to a basis of the habitat $T'_{*}$.

\subsection{ Action of $(\hat{C}_x)^{\delta}$ on $T^{'}_{*}$ }

We now look into the dual action of $(\hat{C}_x)^{\delta}$ on a state $\bra{c',f}\in T^{'}_{*} $. It is sufficient to consider a point $x\in\Sigma$ and the action of $(\hat{C}_x)^{\delta}$:
\begin{eqnarray} \label{cfC}
\bra{c', f}(\hat{C}_x)^{\delta} \equiv \sum_{\phi\in \mathit{Diff}_{c}} f(\phi(x'_1),...,\phi(x'_{n'})) \bra{\, c'\cdot \phi^{-1}}(\hat{C}_x)^{\delta}.
\end{eqnarray}
Now, a term $\bra{c'\circ\phi_1^{-1}}(\hat{C}_x)^{\delta}$ is not zero, only if there exists a charge network $c$, such that $x$ is at one of it's 
non-degenerate vertex, with an edge $I_1$ at $x$ and a color type $j_1$ such that 
\begin{eqnarray}\label{c}
 \ket{c'\circ \phi_1^{-1}} \ =\ \Delta_{[x,I_1,j_1,\delta]}\ket{c}.
\end{eqnarray}
Without loss of generality, we can assume that the vertices of $c_{x,I_1,j_1,\delta}$ are coordinatized as the following. We use $x_1=x, x_2, ... , x_n$ for the vertices shared with $c$. For the newly created vertices, we use $x^{I_1,\delta}$ for the ``apex" vertex and $x^{I_1,\delta,I_1}, ..., x^{I_1,\delta,I_m}$ for the rest, where $I_1,...,I_m$ label edges at $x$. In this notation the corresponding term of (\ref{cfC}) resulting from (\ref{c}) is
\begin{eqnarray} \label{action 0}
f(\phi_1(x'_1),...,\phi_1(x'_{n'})) \bra{\, c'\cdot \phi_1^{-1}}(\hat{C}_x)^{\delta}\ =\ \delta^{-1}   f \left(x,...,x_n, x^{I_1,\delta}, x^{I_1,\delta,I_1}, ..., x^{I_1,\delta,I_m}\right) 
 (s_{x})^{\frac{2}{3}}\cdot q^{j_1}_{I_1}\bra{c} ,
\end{eqnarray}     
where the charged network $c$ is $\delta$ independent and $(s_{x})^{\frac{2}{3}}$ and $q^{j_1}_{I_1}$ refer to the charge network $c$. 
Now, in the result of (\ref{cfC}) this gives the unique term proportional to $\bra{c}$ (if any). That is, we have either 
\begin{eqnarray} \label{action 0}
&&\bra{c', f}(\hat{C}_x)^{\delta} \ket{c}\ =\ \delta^{-1}  f \left(x,...,x_n, x^{I_1,\delta}, x^{I_1,\delta,I_1}, ..., x^{I_1,\delta,I_m}\right) 
 (s_{x})^{\frac{2}{3}}\cdot q^{j_1}_{I_1}
\end{eqnarray}   
when (\ref{c}) holds, or the result of zero if otherwise. The dependence on $\delta$ relevant for the limit in $\delta\rightarrow 0$ is in the factor $$  \delta^{-1}  {f}\left(x,...,x_n, x^{I_1,\delta}, x^{I_1,\delta,I_1}, ..., x^{I_1,\delta,I_m}\right) . $$
At this point we should recall that according to \cite{paper1}, the following holds from the regularization scheme
\begin{eqnarray}
\label{reg feature}
\lim_{\delta\rightarrow 0} \left(x^{I_1,\delta}, x^{I_1,\delta,I_1}, ..., x^{I_1,\delta,I_m}\right) \ &=&\ \left(x, ... , x\right)\nonumber\\ 
\frac{d x^{I_s,\delta}}{d \delta}\ &=&\ \vec{e}_{I_s},\nonumber\\
\frac{d x^{I_s,\delta,I_k},}{d \delta}\ &=&\ 0. 
\end{eqnarray}
This then leads to the result
\begin{eqnarray} \label{action 2}
\delta^{-1}  {f}\left(x,...,x_n, x^{I_1,\delta}, x^{I_1,\delta,I_1}, ..., x^{I_1,\delta,I_m}\right)\
=&&  \delta^{-1} \bigg[ {f}\left( x,...,x_n, x, x, ..., x\right) \nonumber\\ 
&+&\  \delta\, {e}_{I_1}^a\frac{\partial}{\partial y^a}\,{f}\left( x,...,x_n,y,x, ..., x\right)\big|_{ y=x}+O(\delta^2)\bigg]
\end{eqnarray} 

Clearly the limit $\delta\rightarrow 0$ is in general divergent due to the very first term which is of $O(\delta^{-1})$.
Nonetheless, there are special domains in the habitat space $T^{'}_{*}$ on which the limit is well defined.

\subsection{Domains for $\lim_{\delta\rightarrow 0}(\hat{C}_x)^{\delta}$}

One possibility for the domains is to restrict the admissible functions $f$ in the considered states, such that the $O(\delta^{-1})$ term on the right hand side of (\ref{action 2}) is simply zero. As an example, this subspace can be given by 
\begin{eqnarray}
T^{'A}_{*}\equiv Span\{\bra{c',f'}\,;\,{f'}\left(x,...,x_n, x, x, ..., x\right)=0 \}. 
\end{eqnarray} 
On this subspace the limit $\lim_{\delta\rightarrow 0}(\hat{C}_x)^{\delta}$ exists in the point sense, 
\begin{eqnarray} \label{cfC''}
\lim_{\delta\rightarrow 0}\bra{c', f'}(\hat{C}_x)^{\delta} \ket{c}\ =\ {e}_{I_1}^a\frac{\partial}{\partial y^a}\,{f}\left( x,...,x_n,y, x, ..., x\right)\big|_{ y=x}  (s_{x})^{\frac{2}{3}} \cdot q^{j_1}_{I_1}
\end{eqnarray}
when the charge network $c$ satisfies (\ref{c}), or, otherwise
\begin{eqnarray} \label{cfC'}
\lim_{\delta\rightarrow 0}\bra{c', f'}(\hat{C}_x)^{\delta}\ket{c}\ =\  0.
\end{eqnarray}

For our thesis, we now construct a domain relevant for the work by Tomlin and Varadarajan, in a way consistent to their consideration. For simplicity as in their original work, we construct our domain for the quantum scalar constraints using the charge network states with exactly one non-degenerate vertex\footnote{We do not expect difficulties in generalizing the construction upon all the charge networks, although the explicit realization remains to be done.}. Let us go back to (\ref{cfC}) and (\ref{c}), using the notation \eqref{decompdelta} to write 
\begin{eqnarray}
\bra{c',f}\ =\ \bra{c_{x,I_1,j_1,\delta},f},
\end{eqnarray}
and note that this state is actually $\delta$ independent due to (\ref{cphi}). Given the charge network $c$, we consider all the charge networks $c_{x,I_k,j_1,\delta}$, associated with all the edges at the non-degenerate vertex $x$ labeled by $I_1,...,I_m$. The charge type $j_1$ is kept fixed in this set. Also, since there is a one-to-one correspondence between the edges and their deformed images for each $I_k$, the members in $\{c_{x,I_k,j_1,\delta}\}$ enjoy a natural one-to-one correspondences between their sets of vertices, which allows them to share the same function $f$. We then consider a superposition in the form   
\begin{eqnarray}\label{[c,f]}
\bra{c_{x,I_1,j_1,\delta},f}\ +\ \bra{c_{x,I_2,j_1,\delta},f}\ +\ ...\ +\ \bra{c_{x,I_m,j_1,\delta},f}\ =:\ \bra{[c,f,j_1]} . 
\end{eqnarray}
The component of $\bra{[c,f,j_1]}(\hat{C}_x)^{\delta}$ proportional to $\bra{c}$ is then
 \begin{eqnarray} 
 \bra{[c,f,j_1]}(\hat{C}_x)^{\delta}\ \ket{c}\ =\  
\delta^{-1}  \sum_{k=1}^m f\left(x,...,x_n, x^{I_k,\delta}, x^{I_k,\delta,I_1}, ..., x^{I_k,\delta,I_m}\right)  
\cdot q^{j_1}_{I_k} (s_{x})^{\frac{2}{3}} .\end{eqnarray}
Expanding with respect to $\delta$ we obtain
\begin{eqnarray}
 \bra{[c,f,j_1]}(\hat{C}_x)^{\delta} \ \ket{c}\ =\  
&& \delta^{-1} f\left(x,...,x_n, x, x, ..., x\right) (s_{x})^{\frac{2}{3}} \sum_{k=1}^m q^{j_1}_{I_k}  \ +\ 
(s_{x})^{\frac{2}{3}}\sum_{k=1}^m q^{j_1}_{I_k}{e}_{I_k}^a \frac{\partial}{\partial y^a}\,{f}\left( x,...,x_n,  y,x, ..., x\right)\big|_{ y=x} \nonumber\\
&&+\ O(\delta)\ = (s_{x})^{\frac{2}{3}}\sum_{k=1}^m q^{j_1}_{I_k}{e}_{I_k}^a \frac{\partial}{\partial y^a}\,{f}\left( x,...,x_n,  y,x, ..., x\right)\big|_{ y=x}
\ +\ O(\delta)
\end{eqnarray}
where the first term vanished owing to the Gauss constraint (\ref{invariance}).  
Therefore, in the limit
\begin{eqnarray}
\label{dualaction}
\lim_{\delta\rightarrow 0} \bra{[c,f,j_1]}(\hat{C}_x)^{\delta} \ \ket{c}\ =\  
(s_{x})^{\frac{2}{3}}\sum_{k=1}^m q^{j_1}_{I_k}{e}_{I_k}^a \frac{\partial}{\partial y^a}\,{f}\left( x,...,x_n,  y,x, ..., x\right)\big|_{ y=x}.
\end{eqnarray}
In general, the only non-vanishing components of $\lim_{\delta\rightarrow 0} \bra{[c,f]}(\hat{C}_{y})^\delta$ belong to $\{\ket{c\circ\phi^{-1}}; \phi\in \mathit{Diff}_{c}\,,\, \phi(x)=y\}$. The coefficients of these components can be calculated similarly as the above, but with the tangent vectors $\{{e}_{I_k}\}$ of $c$ at $x$ replaced by the corresponding tangent vectors of $c\circ\phi^{-1}$ at $y$. By construction, all the tangent vectors are normalized referring to their associated coordinate patches from the chosen $\{\sigma_{c,i}\}$ (which are just the $\sigma$ for our choice so far). Therefore, the tangent vectors of $c\circ\phi^{-1}$ at $y$ are in the form $\{\lambda(c,I_k,\phi)\cdot \phi_*{e}_{I_k}\}$, where the individual scaling factors $\{\lambda(c,I_k,\phi)\}$ result from the covariance-breaking normalizations. The coefficients are thus given by 
\begin{eqnarray}\label{cfjC}
\lim_{\delta\rightarrow 0} \bra{[c,f,j_1]}(\hat{C}_y)^{\delta} \ \ket{c\circ\phi^{-1}}\ =\  
 (s_{x})^{\frac{2}{3}}\sum_{k=1}^m \lambda(c,I_k,\phi)\,\,q^{j_1}_{I_k}\,\,\phi_*{e}_{I_k}^a\frac{\partial}{\partial y^a}\,f\left( \phi(x),...,\phi(x_n), y,\phi(x), ..., \phi(x)\right)\big|_{y=\phi(x)}.
\end{eqnarray}
In this manner, the operator $\lim_{\delta\rightarrow 0}(\hat{C}_x)^{\delta}$ is well defined on the subspace $ T^{'C}_* \subset T'_*$ given as 
\begin{eqnarray}
T^{'C}_* \equiv Span \{ \bra{[c,f,j]} \}
\end{eqnarray}
Where the $c$ ranges over all the charge networks with one non-degenerate vertex and the $f$ ranges over all the differentiable functions. 

Finally, through our calculation it is clear that the coordinate patches is in the end to determine the tangent vectors appearing in $\hat{C}_y$, thus the only factors in \eqref{cfjC} dependent on $\{\sigma_{c,i}\}$ are the values of $\{\lambda(c,I_k,\phi)\}$ and the specific subset $\mathit{Diff}_{c}$. Using the covariant formulation \cite{paper2} mentioned before, one actually obtains $\lambda(c,I_k,\phi)=1$ with the subset $\mathit{Diff}_{c}$ being the generating set of $\{\sigma_{c,i}\}$, given by $\{\sigma_{c',i} \equiv \sigma_{c,i}\circ\phi^{-1} \,\,;\,\, c'= c\circ\phi^{-1}\}$ for each specific class $[c]$.

\subsection{Solutions to $\hat{C}[N]$ in $T^{'C}_*$}

In order to find the complete solutions of $\hat{C}[N]$ in $T^{'C}_*$, we impose the constraint on a general element in $T^{'C}_*$ through 
\begin{eqnarray}
\label{general solution condition}
\sum_c\sum_{j} \bra{[c,\,f_{(c,j)},\,j]} \hat{C}_y=0.
\end{eqnarray} 
The action of the constraint operators we obtained in the previous chapter implies that the above has to hold for each individual component labeled by one specific $c$, and thus a general solution satisfies
\begin{eqnarray}
\label{condition}
\sum_{j} \bra{[c,\,f_{(c,j)},\,j]} \hat{C}_y=0
\end{eqnarray}
for all $y$. With (\ref{cfjC}) this can be written as
\begin{eqnarray}\label{solution condition} 
 0=\sum_j (s_{x})^{\frac{2}{3}}\sum_{k=1}^m \lambda(c,I_k,\phi)\,\,q^{j}_{I_k}\phi_*{e}_{I_k}^a\frac{\partial}{\partial y^a}\,f_{(c,j)}\left( \phi(x),...,\phi(x_n), y,\phi(x), ..., \phi(x)\right)\big|_{y=\phi(x)}
\end{eqnarray}
for all $y$ and all $\phi\in \mathit{Diff}_{c}$ satisfying $y=\phi(x)$.

 Since we expect the true constraint operator $\hat{C}[N]$, being fully anomaly free, to transform covariantly under the spatial diffeomorphisms, we will now focus on the solutions to \eqref{solution condition} with $\lambda(c,I_k,\phi)=1$. Setting $\lambda(c,I_k,\phi)=1$ in \eqref{solution condition}, we immediately see that the solutions $f_{(c,j)}$ for any $c$ can be characterized according three vectors $\{(\bold{X}_{x,c})_j\}\subset T_x \Sigma $ and their push-forwards $\{\phi_*(\bold{X}_{x,c})_j\}\subset T_y \Sigma$ given by
\begin{eqnarray}
\label{X}
(\bold{X}_{x,c})_j\equiv \sum_{k=1}^m q^{j}_{I_k}{e}_{I_k}\,;\,\,\phi_*(\bold{X}_{x,c})_j = \sum_{k=1}^m q^{j}_{I_k}\phi_*{e}_{I_k}.
\end{eqnarray}

First, we observe that there may be privileged charge network states preferred by $\hat{C}[N]$. Assume that a charge network state $c$ satisfies the condition
\begin{eqnarray}\label{closure}
(\bold{X}_{x,c})_{j_1} =\sum_{k=1}^m q^{j_1}_{I_k}{e}_{I_k}= 0 
\end{eqnarray}
for a specific charge $j_1$. Since this implies $\phi_*(\bold{X}_{x,c})_{j_1}=0$, by setting $f_{(c,j)}=\delta_{j,j_1}\,f$ in (\ref{solution condition}) we immediate see that the resulted state $\bra{[c,f,j_1]}$ with arbitrary $f$ is a solution. Observe that the condition \eqref{closure} characterizing these charge network states takes the form of the Minkowski theorem of polyhedra. In this view, the condition relates the vertex to a polyhedron (labeled by the charge copy $j_1$) in the space of the associated coordinate patch. Specifically, the polyhedron labeled by $j_1$ has the set of faces labeled by $I_k$, with their areas and normal unit vectors (measured in the associated coordinate patch) respectively given by $\{q^{j_1}_{I_k}\}$ and $\{{e}_{I_k}\}$. Due to this, we will call the privileged charge network states in \eqref{closure} the Minkowski states with respect to the charge $j_1$. 

Note that this condition \eqref{closure} is distinct from the various types of ``closure conditions" appearing in canonical loop quantum gravity \cite{closurelqg}, spin foam models\cite{sf}\cite{closuresf} and group field theories\cite{gft}\cite{closuregft}. Particularly, in the canonical loop quantum gravity, the closure condition refers to the Gauss constraints under the interpretation of the Minkowski theorem. Therefore, the closure condition refers to the gauge invariance condition \eqref{invariance} of our model. In this context, the vertex is instead associated to a polyhedron (without the charge labels) in the space spanned by the local triads, having the face areas agreeing with the expectation values of the physical area operators.

Now we can give a complete description to the full solutions to $\hat{C}[N]$ in the distinct subspaces $T^{'C}_*$, each labeled by a certain $c$. In each of these subspaces, the condition (\ref{solution condition}) imposes different restrictions on $f_{(c,j)}$ according to the linear dependency among $\{(\bold{X}_{x,c})_{j_1},(\bold{X}_{x,c})_{j_2},(\bold{X}_{x,c})_{j_3}\}$ as in the following ($\alpha,\beta\in \mathbb{R}$).

(1) The solutions in the subspace labeled by a Minkowski $c$ with respect to all the three charges are spanned by the states labeled by arbitrary functions $(f_{(c,j_1)},f_{(c,j_2)},f_{(c,j_3)})$. 

(2) In the subspace labeled by a $c$ with $(\bold{X}_{x,c})_{j_3}\neq 0$, $(\bold{X}_{x,c})_{j_1}= \alpha (\bold{X}_{x,c})_{j_3}$ and $(\bold{X}_{x,c})_{j_2}= \beta (\bold{X}_{x,c})_{j_3}$, the solutions are spanned by the states labeled by arbitrary $(f_{(c,j_1)},f_{(c,j_2)})$, and the corresponding set of $f_{(c,j_3)}$ satisfying
 \begin{eqnarray}\label{58}
0=\frac{\partial}{\partial y^a}\,\big(\,\alpha f_{(c,j_1)}+\beta f_{(c,j_2)}
+ f_{(c,j_3)}\,\big)\left(x,...,x_n, y,x, ..., x\right)\big|_{y=x}, 
\end{eqnarray}

(3) In the subspace labeled by a $c$ with a linearly independent set $\{(\bold{X}_{x,c})_{j_1},(\bold{X}_{x,c})_{j_2}\}$ and with $(\bold{X}_{x,c})_{j_3}= \alpha (\bold{X}_{x,c})_{j_1} + \beta (\bold{X}_{x,c})_{j_2}$, the solutions are spanned by the states labeled by an arbitrary $f_{(c,j_3)}$, and the corresponding set of $(f_{(c,j_1)},f_{(c,j_2)})$ satisfying 
 \begin{eqnarray}\label{59}
0=\frac{\partial}{\partial y^a}\,\big( f_{(c,j_1)}+\alpha f_{(c,j_3)}\big)\left(x,...,x_n, y,x, ..., x\right)\big|_{y=x} =\frac{\partial}{\partial y^a}\,\big( f_{(c,j_2)}+\beta f_{(c,j_3)}\big)\left(x,...,x_n, y,x, ..., x\right)\big|_{y=x}
\end{eqnarray}

(4) In the subspace labeled by a $c$ with linearly independent $\{(\bold{X}_{x,c})_{j_1},(\bold{X}_{x,c})_{j_2},(\bold{X}_{x,c})_{j_3}\}$, the solutions are spanned by the states labeled by specific $(f_{(c,j_1)},f_{(c,j_2)}, f_{(c,j_3)})$ satisfying
 \begin{eqnarray}\label{60}
0=\frac{\partial}{\partial y^a}\, f_{(c,j_1)}\left(x,...,x_n, y,x, ..., x\right)\big|_{y=x} 
=\frac{\partial}{\partial y^a}\, f_{(c,j_2)}\left(x,...,x_n, y,x, ..., x\right)\big|_{y=x}=\frac{\partial}{\partial y^a}\, f_{(c,j_3)}\left(x,...,x_n, y,x, ..., x\right)\big|_{y=x}\nonumber\\
\end{eqnarray}

Recall that our full system of the quantum constraints consists of not only the newly proposed $\hat{C}[N]$ and $\hat{D}[\vec{\mathcal{N}}_j]$, but also the usual (exponentiated) momentum constraints appearing in the standard loop quantum gravity. Since the usual momentum constraints impose the spatial diffeomorphism symmetry, we now comment on the solutions regarding to such symmetry. First, the relations \eqref{X} hold only under the covariant setting with $\lambda(c,I_k,\phi)=1$, which is the condition for the full constraint system to be anomaly-free. Only in this case, the Minkowski condition \eqref{closure} can be satisfied by all the charge network states in a certain diffeomorphism class $[c]$; when that happens, $[c]$ then yields a Minkowski solution $\bra{[c,f_{(c,j)},j]}$ given an arbitrary $f_{(c,j)}$. However, note that this Minkowski solution itself is not spatial diffeomorphism invariant if the given $f_{(c,j)}$ is not a constant function. On the other hand, we observe that a spatial diffeomorphism invariant state $\bra{[c,f_{(c,j)},j]}$ with any given $[c]$ must have $f_{(c,j)}=\mathit{const}$, and therefore it is automatically a solution to $\hat{C}[N]$.

\section{Action, Domains, and Solution Space of $\hat{D}[\vec{\mathcal{N}}_j]$}

\subsection{The action and the dual action}

The total quantum electric momentum constraint is given by $\hat{D}[\vec{\mathcal{N}}_j]=\sum_{x\in \Sigma}\, N(x) \lim_{\delta \to 0}(\hat{D}^j_x)^\delta$, where the operator $(\hat{D}^j_x)^\delta$ is defined by its action on the charge network basis as
\begin{eqnarray} \label{action 3}
(\hat{D}^j_{x'})^\delta\ket{c} = \sum_{x_i} \,\delta_{x',x_i}\,\,  (s_{x_i})^{\frac{2}{3}} \, \sum_{I_{x_i}} q^j_{I_{x_i}}\, \delta^{-1} \,\left(\Omega_{[x_i,I,\delta]}-\hat{I} \right) \ket{c},
\end{eqnarray}
The part of the sum defined by the term $-\hat{I}$ in the parentheses vanishes again due to the Gauss constraint (\ref{invariance}), therefore
\begin{eqnarray} \label{action 4}
(\hat{D}^j_{x'})^\delta\ket{c} = \sum_{x_i} \,\delta_{x',x_i}\,\, (s_{x_i})^{\frac{2}{3}}\, \sum_{I_{x_i}} q^j_{I_{x_i}}\, \delta^{-1} \,\left(\Omega_{[x_i,I,\delta]}\right)\ket{c}.
\end{eqnarray}

According to \eqref{inner prod}, the general dual action of $\hat{D}^{j,\delta}_{x'}$ removes $\Omega_{[x_i,I,\delta]}$ from a dual state $\bra{c'}$
whenever $c'$ can be written as
\begin{eqnarray}\label{decompdelta2}
c'\ =\Omega_{[x_i,I,\delta]}\cdot c  \equiv\  c_{x_i,I,\delta}, 
\end{eqnarray}
for a certain $x_{i}$ of $c$ with $x_{i}=x'$; otherwise, they simply annihilate $\bra{c'}$. Fortunately, for every charged network $c'$, a decomposition (\ref{decompdelta2}) with any given $(x_{i},I)$ is either unique or absent. Therefore, the operation of $\Omega_{[x_i,I,\delta]}$ is given by 
\begin{eqnarray}
\label{factoring2}
\Omega_{[x_i,I,\delta]}:\bra{c_{x_i,I,\delta}}\ \mapsto\  \bra{c} \,\,\text{, and}\,\,\, \Omega_{[x_i,I,\delta]}:\bra{c'}\ \mapsto \ 0 
\end{eqnarray}
for all other cases.

\subsection{ Action of $(\hat{D}^j_x)^\delta$ on $T^{'}_{*}$ }

 We now look into the dual action of 
$(\hat{D}^j_x)^\delta$ on a state $\bra{c',f}\in T^{'}_{*} $. It is sufficient to consider a point $x\in\Sigma$ and the action of $(\hat{D}^j_x)^\delta$:
\begin{eqnarray} \label{cfC2}
\bra{c', f}(\hat{D}^j_x)^\delta\equiv \sum_{\phi\in \mathit{Diff}_{c}} f(\phi(x'_1),...,\phi(x'_{n'})) \bra{\, c'\circ \phi^{-1}}(\hat{D}^j_x)^\delta.
\end{eqnarray}
Now, a term $\bra{c'\circ\phi_1^{-1}}(\hat{D}^j_x)^\delta$ is not zero, only if there exists a charge network $c$, such that $x$ is it's 
vertex and an edge $I_1$ at $x$, such that 
\begin{eqnarray}\label{c2}
 \ket{c'\circ \phi_1^{-1}} \ =\ \Omega_{[x,I_1,\delta]}\ket{c}.
\end{eqnarray}
We again locate the vertices of $c$ by using $x_1=x, x_2, ... , x_n$. Recall that the action of $\Omega_{[x,I_1,\delta]}$ here removes the non-degenerare node $x_1$. Consistent to the previously chosen notation, we thus denote the vertices in $c'\circ \phi_1^{-1}$ that are shared with $c$ by $x_2, ... , x_n$, and again we denote the remaining outstanding vertices by $x^{I_1,\delta}$ and $x^{I_1,\delta,I_1}, ..., x^{I_1,\delta,I_m}$. Note that the ``skipped vertex"' $x_1=x$ represents the location of this non-trivial action on $c'\circ \phi_1^{-1}$, which is indeed uniquely defined for $c'\circ \phi_1^{-1}$ through the factoring (\ref{c2}).
The corresponding term in (\ref{cfC2}) resulting from (\ref{c2}) is 
\begin{eqnarray} \label{action 0}
f(\phi_1(x'_1),...,\phi_1(x'_{n'})) \bra{\, c'\cdot \phi_1^{-1}}(\hat{D}^j_x)^\delta\ =\ \delta^{-1}   f \left(x_2,...,x_n, x^{I_1,\delta}, x^{I_1,\delta,I_1}, ..., x^{I_1,\delta,I_m}\right) 
 (s_{x})^{\frac{2}{3}}\cdot q^{j}_{I_1}\bra{c} ,
\end{eqnarray}     
where the charged network $c$ is $\delta$ independent and $(s_{x})^{\frac{2}{3}}$ and $q^{j}_{I_1}$ refer to the charge network $c$. 
Now, in the result of (\ref{cfC2}) this gives the unique term proportional to $\bra{c}$ (if any). That is we have either 
\begin{eqnarray} \label{action 0}
\bra{c', f}(\hat{D}^j_x)^\delta\ket{c}\ =\ \delta^{-1}   f \left(x_2,...,x_n, x^{I_1,\delta}, x^{I_1,\delta,I_1}, ..., x^{I_1,\delta,I_m}\right) 
  (s_{x})^{\frac{2}{3}}\cdot q^{j}_{I_1}
\end{eqnarray}   
when (\ref{c2}) holds, or the result of zero if otherwise. The dependence on $\delta$ relevant for the limit in $\delta\rightarrow 0$ is in the factor
$$  \delta^{-1} {f}\left(x_2,...,x_n, x^{I_1,\delta}, x^{I_1,\delta,I_1}, ..., x^{I_1,\delta,I_m}\right) . $$
We again refer to the regularization features in (\ref{reg feature}), and conclude that  
\begin{eqnarray} \label{action 5}
\delta^{-1}  {f}\left(x_2,...,x_n, x^{I_1,\delta}, x^{I_1,\delta,I_1}, ..., x^{I_1,\delta,I_m}\right)\
=&&  \delta^{-1} \bigg[ {f}\left( x_2,...,x_n, x, x, ..., x\right) \nonumber\\ 
&+&\  \delta\, {e}_{I_1}^a\frac{\partial}{\partial y^a}\,{f}\left( x_2,...,x_n,y,x, ..., x\right)\big|_{y=x}+O(\delta^2)\bigg]
\end{eqnarray} 
Clearly the limit $\delta\rightarrow 0$ is in general divergent due to the very first term which is of $O(\delta^{-1})$.
Nonetheless, there are special domains in the habitat space $T^{'}_{*}$, on which the limit is well defined.

\subsection{Domains for $\lim_{\delta\rightarrow 0}(\hat{D}^j_x)^\delta$}

On the domain $T_*^A$ defined previously the limit $\lim_{\delta\rightarrow 0}(\hat{D}^j_x)^\delta$ exists again in the point sense as 
\begin{eqnarray} \label{cfC''}
\lim_{\delta\rightarrow 0}\bra{c', f}(\hat{D}^j_x)^\delta\ket{c}\ =\ {e}_{I_1}^a\frac{\partial}{\partial y^a}\,{f}\left( x_2,...,x_n,  y,x, ..., x\right)\big|_{ y=x} (s_{x})^{\frac{2}{3}}\cdot q^{j_1}_{I_1}
\end{eqnarray}
provided the charge network $c$ satisfies (\ref{c2}), or otherwise
\begin{eqnarray} \label{cfC'}
\lim_{\delta\rightarrow 0}\bra{c', f}(\hat{D}^j_x)^\delta\ket{c}\ =\  0
\end{eqnarray}

Here again we will construct a domain in the spirit of Tomlin and Varadarajan's work. This will be consistent with the above treatment for the scalar constraint, and again we use the charge networks with exactly one non-degenerate vertex. Let us go back to (\ref{cfC2}) and (\ref{c2}), using the notation \eqref{decompdelta2} to write 
\begin{eqnarray}
\bra{c',f}\ =\ \bra{ c_{x,I_{1},\delta},f } ,
\end{eqnarray}
and note that due to (\ref{cphi}) this state is $\delta$ independent. Given the charge network $c$, we consider all the charge networks $ \bra{ c_{x,I_{k},\delta},f } $, associated with all the edges at the non-degenerate vertex $x$ labeled by $I_1,...,I_m$. The members in this set again can naturally share a function, so we can consider a superposition   
\begin{eqnarray}\label{[c,f2]}
\bra{c_{x,I_{1},\delta},f}\ +\ \bra{c_{x,I_{2},\delta},f}\ +\ ...\ +\ \bra{c_{x,I_{m},\delta},f}\ =:\ \bra{[c,f]} . 
\end{eqnarray}
The component of $\bra{[c,f]}(\hat{D}^j_x)^\delta $ proportional to $\bra{c}$ is then
 \begin{eqnarray} 
 \bra{[c,f]}(\hat{D}^j_x)^\delta\ \ket{c}\ =\  
\delta^{-1}  \sum_{k=1}^m f\left(x_2,...,x_n, x^{I_k,\delta}, x^{I_k,\delta,I_1}, ..., x^{I_k,\delta,I_m}\right)  
\cdot q^{j}_{I_k} (s_{x})^{\frac{2}{3}}. \end{eqnarray}
Expanding with respect to $\delta$ we obtain
\begin{eqnarray}
 \bra{[c,f]}(\hat{D}^j_x)^\delta\ \ket{c}\ =\  
&& \delta^{-1} f\left(x_2,...,x_n, x, x, ..., x\right)(s_{x})^{\frac{2}{3}} \sum_{k=1}^m q^{j}_{I_k}  \ +\ 
(s_{x})^{\frac{2}{3}}\sum_{k=1}^m q^{j}_{I_k}{e}_{I_k}^a \frac{\partial}{\partial y^a}\,{f}\left( x_2,...,x_n, y,x, ..., x\right)\big|_{ y=x} \nonumber\\
&&+\ O(\delta)\ = (s_{x})^{\frac{2}{3}} \sum_{k=1}^m q^{j}_{I_k}{e}_{I_k}^a \frac{\partial}{\partial y^a}\,{f}\left( x_2,...,x_n,  y,x, ..., x\right)\big|_{ y=x}
\ +\ O(\delta)
\end{eqnarray}
where the first term vanished owing to the Gauss constraint (\ref{invariance}).  
Therefore, in the limit
\begin{eqnarray}
\label{dualaction2}
\lim_{\delta\rightarrow 0}  \bra{[c,f]}(\hat{D}^j_x)^\delta\ \ket{c}\ =\  
 (s_{x})^{\frac{2}{3}}\sum_{k=1}^m q^{j}_{I_k}{e}_{I_k}^a \frac{\partial}{\partial y^a}\,{f}\left( x_2,...,x_n,  y,x, ..., x\right)\big|_{ y=x}.
\end{eqnarray}
 Similar to the case of the scalar constraints, the only non-vanishing components of $\lim_{\delta\rightarrow 0} \bra{[c,f]}(\hat{D}^j_y)^\delta$ belong to $\{\ket{c\circ\phi^{-1}}; \phi\in \mathit{Diff}_{c}\,,\, \phi(x)=y\}$. The coefficients of these components can be calculated similarly to the above, and using the same scaling factors $\{\lambda(c,I_k,\phi)\}$ introduced before they can be expressed as
 \begin{eqnarray}
\label{cfjC2}
\lim_{\delta\rightarrow 0} \bra{[c,f]} (\hat{D}^j_y)^\delta\ \ket{c\circ\phi^{-1}}\ =\  
 (s_{x})^{\frac{2}{3}}\sum_{k=1}^m \lambda(c,I_k,\phi)\,q^{j_1}_{I_k}\phi_*{e}_{I_k}^a\frac{\partial}{\partial y^a}\,f\left( \phi(x_2),...,\phi(x_n), y,\phi(x), ..., \phi(x)\right)\big|_{y=\phi(x)}.
\end{eqnarray}
In this manner, the operator $\lim_{\delta\rightarrow 0}(\hat{D}^j_x)^\delta$ is well defined in the subspace $ T^{'D}_* \subset T'_*$ defined as
\begin{eqnarray}
T^{'D}_*\equiv \span\{ \bra{[c,f]} \}
\end{eqnarray}
where the $c$ ranges over the charge networks with one non-degenerate vertex and the $f$ ranges over all the differentiable functions.

When the covariant construction is applied, we again have $\lambda(c,I_k,\phi)=1$ using the same subset $\mathit{Diff}_{c}$ that defines $\{\sigma_{c',i}\,\,;\,\,c'\in [c]\}$.

\subsection{Solutions to $\hat{D}[\vec{\mathcal{N}}_j]$ in $T^{'D}_*$ }

In order to find the complete solutions to $\hat{D}[\vec{\mathcal{N}}_j]$ in $T^{'D}_*$, we impose the constraint on a general element in $T^{'D}_*$ through 
\begin{eqnarray}
\label{general solution condition2}
\sum_c \bra{[c,\,f_{c}]}\hat{D}^j_y=0
\end{eqnarray} 
for every $j$. The action of the constraint operators we obtained in the previous chapter implies that the above has to hold for each individual component labeled by one specific $c$, and thus a general solution satisfies
\begin{eqnarray}
\label{condition2}
 \bra{[c,\,f_{c}]} \hat{D}^j_y=0
\end{eqnarray}
for every $j$. Using \eqref{cfjC2}, this condition under the covariant setting can be written as
\begin{eqnarray}\label{solution condition2} 
  0=(s_{x})^{\frac{2}{3}}\sum_{k=1}^m q^{j}_{I_k}\phi_*{e}_{I_k}^a\frac{\partial}{\partial y^a}\,f_{c}\left( \phi(x_2),...,\phi(x_n), y,\phi(x), ..., \phi(x)\right)\big|_{y=\phi(x)=x}
\end{eqnarray}
for every $j$ and $y$, and all $\phi\in \mathit{Diff}_{c}$ satisfying $y=\phi(x)$

Similar to the case of the scalar constraints, for an arbitrary charge network state $c$, the corresponding solutions $f_{c}$ can be characterized by \eqref{solution condition2} using $(\bold{X}_{x,c})_{j}$ defined in \eqref{X}. In this case, there are also states preferred by $\hat{D}[\vec{\mathcal{N}}_j]$ satisfying
\begin{eqnarray}\label{closure2}
(\bold{X}_{x,c})_{j}=\sum_{k=1}^m q^{j}_{I_k}{e}_{I_k}\ =\ 0 
\end{eqnarray} 
for every $j$. With this condition and setting $f_{c}=f$ in \eqref{solution condition2} we immediate see that such a state $\bra{[c,f]}$ with arbitrary $f$ is a solution. That is, the privileged states for $\hat{D}[\vec{\mathcal{N}}_1]$, $\hat{D}[\vec{\mathcal{N}}_2]$ and $\hat{D}[\vec{\mathcal{N}}_3]$ are Minkowski charge network states with respect to all $j$. 

Now we can give a complete our description of the full solutions to $\hat{D}[\vec{\mathcal{N}}_j]$ in the distinct subspaces of $T^{'D}_*$ labeled by the various $c$. In each of these subspaces, the condition (\ref{solution condition2}) imposes different restrictions on $f_{c}$ according to the linear dependency among $\{(\bold{X}_{x,c})_{j_1},(\bold{X}_{x,c})_{j_2},(\bold{X}_{x,c})_{j_3}\}$ as in the following.

(1) The solutions in the subspace labeled by a Minkowski $c$ with respect to all the three charges are spanned by the states labeled by arbitrary functions $f_c$. 

(2) In the subspace labeled by a $c$ with $(\bold{X}_{x,c})_{j_3}\neq 0$ the solutions are spanned by the states labeled by specific $f_c$ satisfying
\begin{eqnarray}\label{61}
0=\frac{\partial}{\partial y^a}\, f_c\left(x_2,...,x_n, y,x, ..., x\right)\big|_{y=x}, 
\end{eqnarray}

Finally, the remarks regarding to the spatial diffeomorphism symmetry in the end of III.E also apply to these solutions.

\section{Solutions in the Anomaly Free Domain $\mathbb D$}

From (\ref{factoring}) and (\ref{factoring2}), it is clear that for any $c$ and the arbitrary $f$ and $f'$, we have $\bra{[c,f,j]}\Omega_{x,I,\delta}=\bra{[c,f']}\Delta_{x,I,j,\delta}=0$ identically with arbitrary values of $\{x,I,j,\delta\}$, and so the domains $T^{'C}_*$ and $T^{'D}_*$ respectively solve the electric momentum and scalar quantum constraints automatically. Combining the two domains, we had identified the full solution subspace ${\mathbb V}$ in $T^{'C}_* \oplus T^{'D}_*$ for both of the constraints given by 
\begin{eqnarray}
\label{solution space}
{\mathbb V}\equiv Span\bigg\{\bra{[\bar{c},f,j]}\,,\, \sum_{j}\bra{[c,\bar{f}_{(c,j)},j]}\bigg\}\oplus Span\bigg\{\bra{[\bar{c},f']}\,,\, \bra{[c,\bar{f}'_c]}\bigg\}.
\end{eqnarray}
Here the $\bar{c}$ ranges over the Minkowski charge network states with respect to all the charges, with $f$ and $f'$ ranging over the arbitrary functions, while the $\bar{f}_{(c,j)}$ and $\bar{f}'_c$ respectively range over the functions satisfying the conditions (1)-(4) in Sec.III.E and the ones satisfying conditions (1) and (2) in Sec.IV.D, with $c$ ranging over arbitrary charge network states (with a single non-degenerate vertex). 

On the other hand, Tomlin and Varadarajan had identified an anomaly free domain $\mathbb D\subset T^{'}_*$ on which the exceptional algebra $$\left[\hat{C}[N],\hat{C}[N']\right]\bigg|_{\mathbb D}=-3\sum_j\left[\hat{D}[\vec{N}_{j}],\hat{D}[\vec{N}'_{j}]\right]\bigg|_{\mathbb D}$$ is nontrivially realized. Combining the two, we expect the space ${\mathbb D}\cap {\mathbb V}$ to contain physical states with the desired semiclassical limits. With our previous analysis, we can now characterize this space.

For simplicity they also generate the basis for $\mathbb D$ from charge network states with a single non-degenerate vertex with $x_1=x$. Because the commutator involves two successive operations of $\Delta_{x,I,j,\delta}$ and  $\Omega_{x,I,\delta}$, the charge network state components involved in $\mathbb D$ are elements from the sets $\{\bra{(c_{x',I',j',\delta'})_{x,I,j,\delta}}\}$ and $\{\bra{(c_{x',I',\delta'})_{x,I,\delta}}\}$ with small enough $\delta$ and $\delta'$. Here the non-degenerate vertex of $c$ is coordinatized by $x'$; the non-degenerate vertex of $c_{x',I',j',\delta'}$ or $c_{x',I',\delta'}$, given by the apex vertex of $\Delta_{x',I',j',\delta'}$ or $\Omega_{x',I',\delta'}$, is coordinatized by $x=x(x',I',\delta')$. Finally, the states from $\{\bra{(c_{x',I',j',\delta'})_{x,I,j,\delta}}\}$ or $\{\bra{(c_{x',I',\delta'})_{x,I,\delta}}\}$ again have one single non-degenerate vertex given by $x^{I,\delta}=x^{I,\delta}(x',I',\delta')$, which is the apex vertex of $\Delta_{x,I,j,\delta}$ and $\Omega_{x,I,\delta}$ with $x=x(x',I',\delta')$. According to (\ref{[c,f]}) and (\ref{[c,f2]}), the states in our domains composed from these elements are given by $$ \bra{[c_{x',I',j',\delta'},f,j]} =\sum_{I}  \bra{(c_{x',I',j',\delta'})_{x,I,j,\delta},f} \,\,\text{and}\,\, \bra{[c_{x',I',\delta'},f']} =\sum_{I}  \bra{(c_{x',I',\delta'})_{x,I,\delta},f'}.$$

To achieve the anomaly freeness, Tomlin and Varadarajan choose a special set of functions $\{f_F,f'_F\}$, consists members defined with a function of a single point $F:\Sigma \to \mathbb{C}$ through  
\begin{eqnarray}
\label{F}
f_F\left(x_1,...,x_n, x^{I,\delta}, x^{I,\delta,I_1}, ..., x^{I,\delta,I_m}\right)\equiv F(x^{I,\delta})\,;\,\,
f'_F\left(x_2,...,x_n, x^{I,\delta}, x^{I,\delta,I_1}, ..., x^{I,\delta,I_m}\right)\equiv F(x^{I,\delta}).
\end{eqnarray}
So these special functions depend only on the location of the non-degenerate vertex of each of the charge network components, which is the final apex vertex obtained after the double actions of the $\Delta$ or $\Omega$ operators on the $c$. The space $\mathbb D$ and is then given by 
\begin{eqnarray}
\label{anomaly free space}
{\mathbb D}\equiv Span\left\{\sum_{j}\sum_{I',j'}\bra{[{c_{x',I',j',\delta'}},f_F,j]}- \frac{1}{12}\sum_{I'}\bra{[{c_{x',I',\delta'}},f'_F]}\right\}
\end{eqnarray}
where $c$ ranges over all charge networks with one nondegenerate vertex, and $F$ ranges over all the differentiable functions.

We may now characterize the solution subspace in $\mathbb D$. Since $\bra{[c_{x',I',j',\delta'},f_F,j]}\in T^{'C}_*$ and $\bra{[c_{x',I',\delta'},f'_F]}\in T^{'D}_*$, we may simply refer to (\ref{solution space}) and conclude that  
\begin{eqnarray}
\label{anomaly free solution space}
{\mathbb D}\cap {\mathbb V} = Span\bigg\{\sum_{j}\sum_{I',j'}\bra{[\overline{{c}_{x',I',j',\delta'}},f_F,j]}- \frac{1}{12}\sum_{I'}\bra{[\overline{{c}_{x',I',\delta'}},f'_F]} \,\,,\nonumber
\\
\sum_{j}\sum_{I',j'}\bra{[{c_{x',I',j',\delta'}},\bar{f}_{\bar{F}},j]}- \frac{1}{12}\sum_{I'}\bra{[{c_{x',I',\delta'}},\bar{f}'_{\bar{F}}]}\bigg\}\,,
\end{eqnarray}
where the set of Minkowski charge network states (with respect to all the charges) must be generated by a certain set of special states $\{c^\star\}$ through $$ \overline{{c}_{x',I',j',\delta'}}\,(c^\star)={c^\star}_{x',I',j',\delta'}\,\,\,\text{and}\,\,\,\overline{{c}_{x',I',\delta'}}\,(c^\star)= {c^\star}_{x',I',\delta'}.$$ Also, the functions $\bar{f}_{\bar{F}}$ and $\bar{f}'_{\bar{F}}$ simply corresponds to $\bar{F}=\mathit{const}$ and reduce to just constants. In conclusion, the space ${\mathbb D}\cap {\mathbb V}$ is given by (\ref{anomaly free solution space}), with the underlying $c^\star $ ranging over the states of one non-degenerate vertex which generate only the Minkowski states (with respect to all the charges), with $F$ ranging over all the differentiable functions, and with $c$ ranging over all charge network states (with a single non-degenerate vertex).

Lastly, based on our previous observations regarding to the spatial diffeomorphism symmetry we comment on this solution space in the anomaly-free domain $\mathbb D$. Again, while the Minkowski solutions in $\mathbb D$ are obtained under the covariant setting, they are not diffeomorphism invariant whenever they involve an $F$ that is not a constant function. On the other hand, any spatial diffeomorphism invariant state in ${\mathbb D}$ must involve only the members in \eqref{anomaly free space} with a constant $F=\bar{F}$, and thus such a state is automatically a solution to the full system of quantum constraints. Note that this points to a potential problem of $\hat{C}[N]$ and $\hat{D}[\vec{N}_{j}]$ being under constraining in the domain $\mathbb D$.

\section{ Summary and Conclusion }

Through our calculations on the quantum constraint system proposed by Tomlin and Varadarajan, we made explicit some important elements inferred by their construction. By writing down the single dual action by the constraint operators $\hat{C}[N]$ and $\hat{D}[\vec{\mathcal{N}}_i]$ on the habitat, we identified a domain $T^{'C}_* \oplus T^{'D}_*$ for both of them. As expected, this domain is much larger than the anomaly-free domain $\mathbb D$ specified by the authors in their original work, which must also support the double actions of the operators. 

In this large domain $T^{'C}_* \oplus T^{'D}_*$, we characterized the the full solution space $\mathbb V$ to $\hat{C}[N]$ and $\hat{D}[\vec{\mathcal{N}}_i]$ under the covariant setting. The space turns out to be spanned by the two classes of solutions-- one with the special underlying charge network states, and the other with the special wave functions. Furthermore, we demonstrated that the two classe of solutions can be naturally projected into $\mathbb D$, thereby leading to the two corresponding classes of solutions spanning the full solution space in $\mathbb D$. The two classes of solutions in $\mathbb V$ are characterized by two remarkable mathematical conditions: the first class of solutions satisfies the condition \eqref{closure} and \eqref{closure2} in the form of the Minkowski theorem of polyhedra, and the second class satisfies a weaker form of the spatial diffeomorphism invariance, as given by \eqref{58}\eqref{59}\eqref{60} and \eqref{61}.

Needless to say, the next step of studying the Tomlin-Varadarajan construction is to check its physical content. In this paper, we have already noticed two potential issues from our preliminary observations on the solution space in $\mathbb D$. The first potential issue is that the Minkowski solutions, having more interesting structures, do not generally satisfy the usual (exponentiated) momentum constraints; the second is that $\hat{C}[N]$ and $\hat{D}[\vec{\mathcal{N}}_i]$ impose no further constraint in the subspace of $\mathbb D$ satisfying the usual momentum constraints. Both of these may be taken as hints that the newly proposed quantum scalar constraints might be incompatible with the usual momentum constraints in loop quantum gravity. In the earlier works\cite{momentumconstr1}\cite{momentumconstr2} by Henderson, Tomlin and Laddha, new quantization schemes similar to the one in our context have been applied to quantize the classical momentum constraints, and they lead to a new type of momentum quantum constraint operators with structures similar to that of $\hat{D}[\vec{\mathcal{N}}_i]$. Our solutions may be suggesting the necessity of replacing the usual momentum constraints completely with this new type of momentum constraints, for a new proper quantum constraint system.

Nevertheless, it is important to point out an alternative scenario in which the two potential issues are absent and our solutions can treated as physical. One may consider applying the construction to the Brown-Kucha\v{r} dust model in the Ashtekar formulation\cite{dust}\cite{dustlqg}. In this case, the model describes gravity coupling to a set of dust scalar fields. Assume the Smolin's limit is still valid, the dust model again has the effective $U^3(1)$ charge symmetry and the same gravitational phase space variables $A_a^i(x)$ and $E^a_i(x)$. The crucial difference is that, with the dust fields serving as the physical spacetime coordinates and their conjugate momenta solving the total constraints, the gravitational sector of the theory is no longer constrained, and the gravitational scalar constraints $C[N]$ become the physical Hamiltonians generating the gravitational evolution with respect to the dust coordinates. Therefore, the Tomlin-Varadarajan quantization procedure may follow as described above, but with the coordinates of $\Sigma$ used in the construction given by the physical dust coordinates. Without the constraints, the domain $\mathbb D$ would be the space of the proper initial physical states, anomaly-free with respect to the evolutions in the dust time generated by $\hat{C}[N]$. Furthermore, in this scenario the solution space ${\mathbb D}\cap {\mathbb V}$ would actually represent the space of physical ground states.

Certainly, a more detailed study is still needed to decide whether these potential issues are real, or whether the solution space can contain proper physical degrees of freedom. Moreover, it remains to be seen if the solution space can carry an inner product for a physical Hilbert space to be defined. Since our solution space is explicitly given, these questions can now be explicitly answered for the examination of the model's validity.

\section{ Acknowledgements }
We benefited a lot from numerous discussions with Alok Laddha, Casey Tomlin and Madhavan Varadarajan as well as with Eugenio Bianchi and Wolfgang Wieland. This work was supported by the grant of the Polish Narodowe Centrum Nauki nr. 2011/02/A/ST2/00300.


\end{document}